# AN Alternative Two-Fluid Formulation Of a Partially Ionized Plasma

V. Krishan

Indian Institute of Astrophysics, Bangalore-560034, India

E-mail: vinod@iiap.res.in

**Abstract**

In a recent paper (Krishan 2021), a two - fluid description of a partially ionized plasma was presented in which electron fluid and the neutral fluid were combined appropriately into one fluid, christened as ENe fluid, and treat the ions as the second fluid. Some of the electrostatic modes of this two-fluid system were studied. Here, I discuss another possibility in which the ion fluid and the neutral fluid are combined into one fluid, christened as INe fluid, and treat the electrons as the second fluid. There can be a huge variation between the relative masses of the neutrals and the ions. Thus both have to be treated as the inertia carrying species. After establishing the framework for the INe-electron fluids, some of the characteristic wave modes of this novel plasma are investigated.

## 1. Introduction

A multi-species particle system is hard to crack. Often, simplifications are sought which can reduce the effective degrees of freedom of the system. Statistical averaging is a powerful technique by which a collection of a large number of particles can be transformed into a single fluid with its characteristic properties of mass density, flow velocity, pressure and viscosity. It is like getting water from water molecules. The simplest plasma is a fully ionized plasma consisting of electrons and ions. This plasma can be described by a single conducting fluid by appropriately combining the dynamics of the electrons and the ions Alfve`n and Falthammer (1957). The magnetohydrodynamic (MHD) fluid description so obtained has served well in accounting for a variety of phenomena in the area of plasma physics. The simplest partially ionized plasma consists of three species of particles viz. electrons, ions and the neutrals. Such a system can be modeled in diverse ways. One can describe it as a three-fluid system consisting of an electron fluid, an ion fluid and a neutral fluid. The three fluids can again be appropriately combined into a single conducting fluid in the manner of the MHD ( Krishan, V. 2014 and references therein). This system has been investigated variously, e.g. Krishan and Gangadhara, (2008); Hiryaki et al. (2010); Krishan and Varghese (2009), Ganadhara et al (2014). , Paradhkar et al (2019).

If one desires more information about the system, the two-fluid description can be accessed in three ways 1) one fluid by combining the electron and the ion fluids to obtain the MHD conducting fluid and the second the neutral fluid, 2) one fluid by combining the electron and the neutral fluids and the second the ion fluid and 3) one fluid by combining the ion and the neutral fluids and the second the electron fluid. While there is some work on the the first possibility, the second possibility, to the best of my knowledge, have been explored only recently (Krishan, 2021). In this paper I explore the third possibility. Here, in section 2, the dynamics of the ion and the neutral fluids will be combined to generate a single fluid which is christened as the INe fluid. In section 3, the dynamics of the electron fluid is described. The Poisson equation is established in section 4. The study of some of the normal modes in the two, INe and electron fluids is presented in sections 5-7. The paper is concluded in section 8.

## 2 . The INe Fluid

The INe fluid is generated by combining the ion and the neutral fluids appropriately in the manner of the MHD. We derive the mass, momentum , the electric charge and the energy conservation laws for the INe fluid.

We begin with the mass conservation law

$$\frac{\partial \rho_i}{\partial t} + \nabla \cdot (\rho_i V_i) = 0 \qquad (1)$$

of the ion fluid of mass density $\rho_i$ and velocity $V_i$.

The mass conservation law of the neutral fluid of mass density $\rho_n$ and velocity $V_n$ is

$$\frac{\partial \rho_n}{\partial t} + \nabla \cdot (\rho_n V_n) = 0 \quad (2)$$

Adding Eqs. (1) and (2) begets the mass conservation law of the INe fluid

$$\frac{\partial \rho}{\partial t} + \nabla \cdot (\rho V) = 0 \quad (3)$$

where

$$\rho = \rho_i + \rho_n$$

is the mass density and

$$V = \frac{\rho_i V_i + \rho_n V_n}{\rho}$$

is the center of mass velocity of the INe fluid.

The momentum conservation law for the neutral fluid is

$$\rho_n D_n = -\nabla P_n - \rho_n \nu_{ne}(V_n - V_e) - \rho_n \nu_{ni}(V_n - V_i) \quad (4)$$

where

$$D_n = [\frac{\partial V_n}{\partial t} + (V_n \cdot \nabla) V_n]$$

The momentum conservation law for the ion fluid is

$$\rho_i D_i = en_i\left[E + \frac{V_i \times B}{c}\right] - \nabla P_i - \rho_i \nu_{ie}(V_i - V_e) - \rho_i \nu_{in}(V_i - V_n) \quad (5)$$

where

$$D_i = [\frac{\partial V_i}{\partial t} + (V_i \cdot \nabla) V_i]$$

On defining the relative velocity $u$ to be

$$u = (V_i - V_n) \quad (6)$$

we find

$$V_i = V + \frac{\rho_n}{\rho} u,$$

$$V_n = V - \frac{\rho_i}{\rho} u \quad (7)$$

We can now substitute for $V_i$ and $V_n$ In Eqs. (4) and (5) from Eq.(7) and add them up. Neglecting the nonlinear convective derivative terms as is done in the formulation of MHD, we find

$$\rho \frac{\partial V}{\partial t} + \rho_i u \frac{\partial}{\partial t} \ln(\frac{\rho}{\rho_i}) = qn\left[E + \frac{(V + \frac{\rho_n}{\rho} u) \times B}{c}\right] - \nabla P$$

$$-\nabla \rho_n (\frac{\rho_i}{\rho}) u (\nu_{ie} - \nu_{ne}) \quad (8)$$

where

$$P = P_i + P_n, \quad qn = en_i \quad (9)$$

Eq (8) describes the momentum conservation law of the new INe fluid with number density $n$, mass density $\rho$, velocity $V$, pressure $P$ and an effective charge density $(qn) = (en_i)$. The INe fluid is thus seen to consists of particles of mass $m = \rho/n$, electric charge $q=$ ($en_i/n$). The additional terms depending upon the ion and electron fluid characteristics will be considered as the nonideal effects.

The difference of Eqs.(4) and (5) gives the evolution of the relative velocity as

$$\frac{\partial u}{\partial t} = \frac{e}{m_i}\left[E + \frac{(V + \frac{\rho_n}{\rho}u) \times B}{c}\right] - \frac{\nabla P_i}{\rho_i} - \frac{\nabla P_n}{\rho_n} - u(\nu_{in} + \nu_{ni}) \quad (10)$$

in the limit $\rho_e \to 0$.

We should also address the question of electric charge conservation. The ion fluid charge conservation is described as

$$\frac{\partial(en_i)}{\partial t} + \nabla \cdot (en_i V_i) = 0 \quad (11)$$

which can be written in terms of the charge density of the INe fluid as

$$\frac{\partial(qn)}{\partial t} + \nabla \cdot [qn(V + V_i - V)] = 0 \quad (12)$$

On using mass conservation, Eq.(3) and substituting for the velocity difference $(V_i - V)$ from Eq. (7), we find

$$n[\frac{\partial q}{\partial t} + V \cdot \nabla q] + \nabla \cdot [nq\frac{\rho_n}{\rho}u] = 0 \quad (13)$$

Or

$$n\frac{dq}{dt} + \nabla \cdot [nq(1 - \frac{\rho_i}{\rho})u] = 0 \quad (14)$$

This along with Eq.(10) for $u$ determines the charge density variation of the INe fluid. One can see that in the equilibrium with no relative flow between the neutral and the ion fluids i.e. for $u = 0$, $q = q_0 = (en_{i0}/n_0)$.

The energy conservation for the INe fluid is described by taking the equation of state to be

$$\nabla P = c^2 \nabla \rho \quad (15)$$

where $P, \rho$ and $c^2$ are respectively the pressure, the density and the square of the sound speed of a fluid.

Thus the mass, momentum, electric charge density and the energy conservation of the INe fluid are respectively contained in Eqs.(3), (8), (14) and (15).

**3. The Electron Fluid**

The mass conservation for the electron fluid is

$$\frac{\partial \rho_e}{\partial t} + \nabla \cdot (\rho_e V_e) = 0 \quad (16)$$

The momentum conservation of the electron fluid is :

$$\rho_e \left[ \frac{\partial V_e}{\partial t} + (V_e . \nabla) V_e \right] =$$

$$-e n_e \left[ E + \frac{V_e \times B}{c} \right] - \nabla P_e - \rho_e$$

$$\nu_{en}(V_e - V_n) - \rho_e \nu_{ei}(V_e - V_i) \quad (17)$$

where $\nu_{en}$ and $\nu_{ei}$ are the electron-neutral and the electron-ion collision frequencies. We can substitute for the neutral and the ion velocities in terms of the INe fluid velocity $V$ and the relative velocity u to get

$$\left[ \frac{\partial V_e}{\partial t} + (V_e . \nabla) V_e \right] = - \frac{e n_e}{\rho_e} \left[ E + \frac{V_e \times B}{c} \right]$$

$$- \nabla P_e - \nu_{en} \left( V_e - V + \frac{\rho_i}{\rho} u \right) -$$

$$\nu_{ei} \left( V_e - V - \frac{\rho_n}{\rho} u \right) \quad (18)$$

The charge conservation for the electron fluid is

$$\frac{\partial (e n_e)}{\partial t} + \nabla . (e n_e V_e) = 0 \quad (19)$$

The equation of state for the electron fluid is

$$\nabla P_e = c_e^2 \nabla \rho_e \quad (20)$$

where $c_e^2$ is the square of the sound speed in the electron fluid.

Thus Thus the mass, momentum, electric charge density and the energy conservation of the electron fluid are respectively contained in Eqs.(16), (18), (19) and (20).

**4 . The Poisson Equation**

The electric field is described through the Poisson equation as:

$$\nabla . E = 4\pi (e n_i - e n_e) \quad (21)$$

The electromagnetic waves in magnetized INe and electron fluids can be studied by using the wave equation

$$\nabla^2 E - \nabla (\nabla . E) = \frac{4\pi}{c^2} \frac{\partial J}{\partial t} + \frac{1}{c^2} \frac{\partial^2 E}{\partial t^2}$$

(22)

where J , the current density, is defined as

$$J = e(n_i V_i - n_e V_e) \quad (23)$$

which in terms of the INe fluid velocities becomes

$$J = e[n_i (V + \frac{\rho_n}{\rho} u) - n_e V_e] \quad (24)$$

We now have the complete mathematical formulation of the dynamics of the two fluids, the INe and the electron fluids along with the Poisson equation and the electromagnetic wave equation to study the existence of electrostatic and electromagnetic waves.

**5. The Linear Electrostatic Waves**

The equilibria of the INe and the electron fluids are taken to be uniform and static. After substituting the one dimensional plane wave variation, exp($ikx-i\omega t$), of all the perturbed quantities in the linearized form of Eqs. (3), (9), (10), in the absence of a magnetic field, one gets

$$-\omega\rho_1 + k\rho_0 V_1 = 0 \quad (3a)$$

$$\omega\rho_0 V_1 = iqn_o E_1 - ia\, u_1 + kP_1 \quad (8a)$$

$$P_1 = c^2 \rho_1 \quad (15a)$$

$$(\omega + i\nu_i)u_1 = \frac{ieE_1}{m_i} + \frac{kP_i}{\rho_{i0}} - \frac{kP_n}{\rho_{n0}} \quad (10a)$$

$$-\omega\rho_{e1} + k\rho_{e0}V_{e1} = 0 \quad (16a)$$

$$(\omega + i\nu_e)V_{e1} = -\frac{ieE_1}{m_e} + i\nu_e V_1 + ibu_1 + \frac{kP_{e1}}{\rho_{e0}} \quad (18a)$$

$$P_{e1} = c_e^2 \rho_{e1} \quad (20a)$$

$$ikE_1 = 4\pi\, e(n_{i1} - n_{e1}) \quad (21a)$$

where

$$a = \rho_{n0}\left(\frac{\rho_{i0}}{\rho_0}\right)(\nu_{ie} - \nu_{ne})$$

$$= \left(\frac{\rho_{e0}}{\rho_0}\right)(\nu_{ei}\rho_{n0} - \nu_{en}\rho_{i0}) \to 0,$$

for $\rho_{e0} \to 0$ (25)

and $\quad b = \dfrac{a}{\rho_{e0}}$

Substituting for the density perturbations from the continuity equations, the Poisson Equation becomes

$$iE_1 = \frac{4\pi e n_{i0}}{\omega}\left(V_1 + \left(\frac{\rho_{n0}}{\rho_0}\right)u_1 - V_{e1}\right) \quad (26)$$

The dispersion relation of the electrostatic waves can be determined using the well known procedure of substituting for all the perturbed quantities in terms of one of them. This gives

1. for all pressure perturbations and collision frequencies = 0,

$$V_1 = \frac{ien_{i0}E_1}{\omega\rho_0}, \quad u_1 = \frac{ieE_1}{\omega m_i},$$

$$V_{e1} = \frac{ieE_1}{\omega m_e} \quad (27)$$

and the dispersion relation is

$$\omega^2 = \omega_{ep}^2 + \omega_{ip}^2 \quad (28)$$

Thus the role of the neutrals disappears in the absence of collisions and pressure perturbations.

2. retaining pressure perturbations $P_1$ in the INe fluid and $P_{e1}$ in the electron fluid with all collision frequencies = 0, one gets

$$V_1 = \frac{ien_{i0}E_1}{\omega\rho_0}(1-\frac{k^2c^2}{\omega^2})^{-1},$$

$$u_1 = \frac{ieE_1}{\omega m_i},$$

$$V_{e1} = \frac{ieE_1}{\omega m_e}(1-\frac{k^2c_e^2}{\omega^2})^{-1} \quad (29)$$

The dispersion relation is now found to be

$$1 - (\frac{\rho_{i0}}{\rho_0})\frac{\omega_{ip}^2}{\omega^2 - k^2c^2} - (\frac{\rho_{n0}}{\rho_0})\frac{\omega_{ip}^2}{\omega^2} - \frac{\omega_{ep}^2}{\omega^2 - k^2c_e^2} = 0 \quad (30)$$

In the limit $k^2c^2$ and $k^2c_e^2$ much less than $\omega^2$, one recovers the dispersion relation, Eq. (28). In the opposite limit when $k^2c^2$ and $k^2c_e^2$ are both much larger than $\omega^2$, one gets

$$\omega^2 = (\frac{\rho_{n0}}{\rho_0})\omega_{ip}^2$$

$$[1+(\frac{\rho_{i0}}{\rho_0})\frac{\omega_{ip}^2}{k^2c^2}+\frac{\omega_{ep}^2}{k^2c_e^2}]^{-1}$$

(31)

which can also be written in terms of the plasma frequency of the INe fluid as

$$\omega^2 = (\frac{\rho_{n0}}{\rho_{i0}})\omega_p^2[1+\frac{\omega_p^2}{k^2c^2}+\frac{\omega_{ep}^2}{k^2c_e^2}]^{-1} \quad (32)$$

where

$$\omega_p^2 = \frac{4\pi n_0(\frac{n_{i0}e}{n_0})^2}{m}$$

(33)

is the plasma frequency of the INe fluid consisting of particles of mass $m$, equilibrium density $n_0$ and equilibrium electric charge

$$(\frac{n_{i0}e}{n_0}).$$

The dispersion relation in Eq. (32) may be christened as the INe-acoustic mode since it is reminiscent of the ion-acoustic wave dispersion in a fully ionized plasma. One may note that the INe-acoustic mode vanishes in a fully ionized plasma.

3. Retaining all collision frequencies and neglecting all pressure perturbations, from Eqs. (8a) and (10a), one gets,

$$V_1 = i\frac{en_{io}E_1}{\omega\rho_0} \quad (34)$$

$$u_1 = i\frac{eE_1}{m_i(\omega + i\nu_i)} \quad (35)$$

From Eq. (18a), with the neglect of all pressure perturbations, the electron fluid momentum equation gives

$$(\omega + i\nu_e)V_{e1} = -\frac{ieE_1}{m_e} + i\nu_e V_1 + ibu_1 \quad (36)$$

After substituting for $V_1$ and $u_1$, one gets

$$V_{e1} = -\frac{ieE_1}{m_e}\left[\frac{\omega\rho_0 - i\nu_e\rho_{e0}}{\omega\rho_0(\omega + i\nu_e)} - i\frac{bm_e}{m_i(\omega + i\nu_e)(\omega + i\nu_i)}\right] \quad (37)$$

After substituting for the velocities in the Poisson, Eq.(21), the dispersion relation of the electrostatic waves can be determined. In the low frequency limit such that $\omega \ll \nu_e$ and $\nu_i$ we find

$$\omega = -i\left[\frac{\omega_{ip}^2}{\nu_i} + \frac{\omega_{ep}^2}{\nu_e} + \frac{b\omega_{ip}^2}{\nu_e\nu_i}\right] \quad (38)$$

which is a purely damped mode for $b > 0$. Now from Eq.(22) $b$ can be shown to be

$$b = \left(\frac{1}{\rho_0}\right)(\nu_{ei}\rho_{n0} - \nu_{en}\rho_{i0}) =$$

$$\nu_{ei} - (\nu_{ei}+\nu_{en})\frac{\rho_{i0}}{\rho_0} \quad (39)$$

(39)

which could acquire a negative value under special conditions. Nevertheless the other two terms in Eq.(38) are larger than the b term and thus the mode remains damped.

One could also consider the case for $\nu_e \gg \omega \gg \nu_i$ and the dispersion relation turns out to be

$$\omega^2 + \frac{i\omega_{ep}^2\omega}{\nu_e} - \omega_p^2\frac{\rho_0}{\rho_{i0}} = 0 \quad (40)$$

where $b \approx -\nu_e\frac{\rho_{i0}}{\rho_0}$

has been used.

## 6. Electrostatic Waves in Magnetized INe and Electron Fluids

We determine the dispersion relation of electrostatic waves in the presence of a uniform magnetic field $B_0$, say in the z direction. The electrostatic waves have their electric field $E_{1x}$ and the wave vector k parallel to each other and perpendicular to the uniform magnetic field.

The linearized form of the momentum equation (8) of the INe fluid with the plane wave variation for the first order quantities reads

$$V_{1x} = \frac{i\omega_{ec}\rho_{e0}}{\omega\rho_0}\left(V_{1y} + \frac{\rho_{n0}}{\rho_0}u_{1y}\right) + \frac{ien_{i0}E_{1x}}{\omega\rho_0} \quad (8b)$$

$$V_{1y} = -\frac{i\omega_{ec}\rho_{e0}}{\omega\rho_0}(V_{1x} + \frac{\rho_{n0}}{\rho_0}u_{1x})$$

$$V_{1z} = 0$$

which furnish

$$V_{1x} = \frac{ieE_{1x}}{\omega m}\frac{T}{(1-S)} \quad (41)$$

where

$$T = \frac{n_{i0}}{n_0}(1-\frac{\alpha_i\omega_{ic}^2}{\omega^2})^{-1}(1-\frac{\alpha_n\omega_{ic}^2}{\omega^2})^{-1}$$

$$S = \frac{\alpha_i\omega_{ic}^2}{\omega^2}\frac{\alpha_i\omega_{ic}^2}{\omega^2}(1-\frac{\alpha_i\omega_{ic}^2}{\omega^2})^{-1}(1-\frac{\alpha_n\omega_{ic}^2}{\omega^2})^{-1} \quad (42)$$

$$\alpha_n = \frac{\rho_{n0}}{\rho_0}, \quad \alpha_i = \frac{\rho_{i0}}{\rho_0}$$

The corresponding electron fluid equation gives

$$V_{ex} = -\frac{ieE_{1x}}{\omega m_e}(1-\frac{\omega_{ec}^2}{\omega^2})^{-1} \quad (18b)$$

Equation (15) for the relative velocity between ions and neutrals gives

$$u_{1x} = \frac{ieE_{1x}}{\omega m}\frac{1}{(1-\frac{\rho_{n0}}{\rho_0}R)}(\frac{RT}{(1-S)} + \frac{m}{m_i}) \quad (43)$$

where

$$R = \frac{\omega_{ic}^2}{\omega^2} \quad (44)$$

Substitution in the Poisson equation

$$1 = \frac{4\pi e n_{i0}}{\omega}(V_{1x}` + (\frac{\rho_{n0}}{\rho_0})u_{1x} - V_{e1x}) \quad (45)$$

begets the dispersion relation of the electrostatic waves in the presence of the uniform magnetic field

$$1 - \frac{\omega_{ip}^2}{\omega^2}(1-\frac{\alpha_n\omega_{ic}^2}{\omega^2})^{-1}[\alpha_n\frac{\omega_{ic}^2}{\omega^2} + \frac{\alpha_i}{(1-\frac{\alpha_n\omega_{ic}^2}{\omega^2})(1-\frac{\alpha_i\omega_{ic}^2}{\omega^2}) - \alpha_n\alpha_i\frac{\omega_{ic}^4}{\omega^4}}] - \frac{\omega_{ep}^2}{\omega^2}(1-\frac{\omega_{ec}^2}{\omega^2})^{-1} = 0 \quad (46)$$

For $\alpha_n = 0$ so that $\alpha_i = 1$ and stationary ions, one can easily recover the dispersion relation

$$\omega^2 = \omega_{ep}^2 + \omega_{ec}^2 \quad (47)$$

of the upper hybrid waves in an electron-ion plasma. For $\alpha_n = 0$ and in the plasma approximation, $n_{e1} = n_{i1}$, amounting to the neglect of first term one recovers the dispersion relation

$$\omega^2 = \omega_{ic}\omega_{ec} \quad (48)$$

of the lower hybrid waves in an electron-ion plasma. In order to determine the effect of the degree of ionization on these modes, the dispersion relation, Eq.(46) must be plotted as a function of $\alpha_n$ or $\alpha_i$. the inclusion of collisional terms contribute to the damping of the wave modes.

## 7. Electromagnetic Waves in Magnetized INe and Electron Fluids

The electromagnetic waves in magnetized INe and Electron Fluids can be studied by using equations (22 and (24).

The linearization gives

$$J_1 = en_{i0}[V_1 + \alpha_n u_1 - V_{e1}] \quad (49)$$

where in the equilibrium, $n_{i0} = n_{e0}$ has been used. We can study the electromagnetic waves propagating parallel and perpendicular to the ambient magnetic field.

### 7a. Ordinary Wave

We first consider the perpendicular propagation such that $B_0 = B_z$, $k = k_x$ and the electric field, $E_1 = E_{1z}$, is parallel to $B_0$. This characterizes the ordinary wave. The linearized wave equation becomes

$$(\omega^2 - k^2 c^2)E_{1z} = -4\pi ei\omega n_{e0}[V_{1z} + \alpha_n u_{1z} - V_{e1z}] \quad (50)$$

On examining the z components of the velocities, we find that the predominant contribution is from the electron fluid. The other two terms contribute terms of the order or smaller than the ion plasma frequency. Substituting for $V_{e1z}$ fetches

$$\omega^2 = \omega_{ep}^2(1 + i\frac{\nu_e}{\omega})^{-1} + k^2 c^2 \quad (51)$$

It shows that the ordinary wave undergoes damping predominantly due to electron-neutral collisions.

### 7b. Extraordinary wave

This wave is characterized by $k=k_x$, $E_1 = (E_{1x}, E_{1y})$ which is perpendicular to $B_0$. The wave equation takes the form

$$\omega^2 E_{1x} = -4\pi ei\omega n_{e0}[V_{1x} + \alpha_n u_{1x} - V_{e1x}]$$

$$(\omega^2 - k^2 c^2)E_{1y} = -4\pi ei\omega n_{e0}[V_{1y} + \alpha_n u_{1y} - V_{e1y}] \quad (52)$$

The extraordinary wave is also a high frequency wave. Therefore the electron fluid velocity is the main contributor to the current density. The INe fluid contributes terms of the order or less than the ion plasma frequency. One can determine the dispersion relation including the electron- neutral collision frequency.

## 8. Conclusion

The new fluid, here, christened as the INe fluid is obtained by combining the ion and the neutral fluids appropriately. A partially ionized plasma thus can be described by two fluids the INe fluid and the electron fluid. Some of the normal modes in this system have been investigated. The role of the degree of ionization is obvious. A more detailed parametric study of the various modes is highly desirable and would bring out the novelties of the system.

**Acknowledgements**

Discussions with Professor Abhijit Sen are gratefully acknowledged.